\documentclass[twocolumn,amsmath,aps,prd,preprintnumbers,amssymb,nofootinbib,showpacs,floatfix]{revtex4}
\usepackage{amssymb}
\usepackage{amsmath}
\usepackage{epsfig}
\usepackage{subfigure}
\usepackage{mathrsfs}
\usepackage{longtable}
\usepackage[usenames,dvipsnames]{xcolor}
\usepackage{mathtools}
\usepackage{relsize}
\begin{document}
%Dan's definitions
 \renewcommand{\thefigure}{\arabic{figure}}
\newcommand{\noj}{}

\newcommand{\apjl}{Astrophys. J. Lett.}
\newcommand{\aap}{Astron. Astrophys.}
\newcommand{\apjs}{Astrophys. J. Suppl. Ser.}
\newcommand{\sa}{Sov. Astron. Lett.}.
\newcommand{\jpb}{J. Phys. B.}
\newcommand{\natu}{Nature (London)}
\newcommand{\aaps}{Astron. Astrophys. Supp. Ser.}
\newcommand{\aj}{Astron. J.}
\newcommand{\aas}{Bull. Am. Astron. Soc.}
\newcommand{\mnras}{Mon. Not. R. Astron. Soc.}
\newcommand{\pasp}{Publ. Astron. Soc. Pac.}
\newcommand{\jcap}{JCAP.}
\newcommand{\jmat}{J. Math. Phys.}
\newcommand{\prep}{Phys. Rep.}
\newcommand{\jtep}{Sov. Phys. JETP.}
\newcommand{\plb}{Phys. Lett. B.}
\newcommand{\pla}{Phys. Lett. A.}
\newcommand{\jhep}{Journal of High Energy Physics}

%Doddy's definitions
\newcommand{\be}{\begin{equation}}
\newcommand{\ee}{\end{equation}}
\newcommand{\bea}{\begin{align}}
\newcommand{\eea}{\end{align}}
\newcommand{\lsim}{\mathrel{\hbox{\rlap{\lower.55ex\hbox{$\sim$}} \kern-.3em \raise.4ex \hbox{$<$}}}}
\newcommand{\gsim}{\mathrel{\hbox{\rlap{\lower.55ex\hbox{$\sim$}} \kern-.3em \raise.4ex \hbox{$>$}}}}
\newcommand{\grad}{\ensuremath{\vec{\nabla}}}
\newcommand{\adotoa}{\ensuremath{{\cal H}}} 
\newcommand{\Uc}{\ensuremath{{\cal U}}}
\newcommand{\Vc}{\ensuremath{{\cal V}}}
\newcommand{\Jc}{\ensuremath{{\cal J}}}
\newcommand{\Mc}{\ensuremath{{\cal M}}}

\newcommand{\unit}[1]{\ensuremath{\, \mathrm{#1}}}

%Ewan's definitions
\newcommand{\gb}{\gamma_{\rm b}}
\newcommand{\dx}{\delta x}
\newcommand{\dy}{\delta y}
\newcommand{\dz}{\delta z}
\newcommand{\dr}{\delta r}
\newcommand{\ds}{\delta s}
\newcommand{\dt}{\delta t}
\newcommand{\uns}{\rmunderscore}
\newcommand{\chimin}{\langle \chi \rangle}

%Scott's definitions
\newcommand{\bi}{\begin{itemize}}
\newcommand{\ei}{\end{itemize}}
\newcommand{\ben}{\begin{enumerate}[itemsep=0.1in,parsep=0.1in]}
\newcommand{\een}{\end{enumerate}}
\newcommand{\tben}{\begin{enumerate}[itemsep=0.0in,parsep=0.0in]}
\newcommand{\teen}{\end{enumerate}}
\newcommand{\ud}{{\rm d}}
\newcommand{\drm}{\mathrm{d}}
\newcommand{\rhos}{\rho_\phi}
\newcommand{\rhor}{\rho_r}
\newcommand{\pr}{p_r}
\newcommand{\rhom}{\rho_\mathrm{dm}}
\newcommand{\rhost}{\tilde{\rho}_\phi}
\newcommand{\gam}{\Gamma_\phi}
\newcommand{\cms}{\; {\rm cm^3}/{\rm s}}
\newcommand{\gev}{\; \mbox{GeV}}
\newcommand{\tev}{\; \mbox{TeV}}

%%%%%%%%%%%%% colortext comments %%%%%%%%%%%%%%
\newcommand{\tkDM}[1]{\textcolor{red}{#1}}                     % Doddy
\newcommand{\tkGSW}[1]{\textcolor{blue}{#1}}		  % Scott
%%%%%%%%%%%%%%%%%%%%%%%%%%%%%%%%%%%%%%%%%%

%\title{The Implications of Tensor Detection for Axion Dark Matter}
\title{Tensor Detection Severely Constrains Axion Dark Matter}

\author{David J. E. Marsh$^{1}\footnote{dmarsh@perimeterinstitute.ca} 
$,~Daniel Grin$^{2}$,~Ren\'{e}e Hlozek$^{3}$, and Pedro G. Ferreira$^{4}$}
\affiliation{$^{1}$Perimeter Institute, 31 Caroline St N,  Waterloo, ON, N2L 6B9, Canada}
\affiliation{$^{2}$Department of Astronomy and Astrophysics, University of Chicago, Illinois, 60637, U.S.A. }
     \affiliation{$^{3}$Department of Astronomy, Princeton University, Princeton, NJ 08544, USA}
 \affiliation{$^{4}$Astrophysics, University of Oxford, DWB, Keble Road, Oxford, OX1 3RH, UK}

\date{\today}

 %---------------------- ABSTRACT -------------------------
\begin{abstract}

The recent detection of B-modes by BICEP2 has non-trivial implications for axion dark matter implied by combining the tensor interpretation with isocurvature constraints from Planck. In this paper the measurement is taken as fact, and its implications considered, though further experimental verification is required. In the simplest inflation models $r=0.2$ implies $H_I=1.1\times 10^{14}\text{ GeV}$. If the axion decay constant $f_a<H_I/2\pi$ constraints on the dark matter (DM) abundance alone rule out the QCD axion as DM for $m_a \lesssim 52\chi^{6/7}\,\mu\text{eV}$ (where $\chi>1$ accounts for theoretical uncertainty). If $f_a>H_I/2\pi$ then vacuum fluctuations of the axion field place conflicting demands on axion DM: isocurvature constraints require a DM abundance which is too small to be reached when the back reaction of fluctuations is included. High $f_a$ QCD axions are thus ruled out. Constraints on axion-like particles, as a function of their mass and DM fraction, are also considered. For heavy axions with $m_a\gtrsim 10^{-22}\text{ eV}$ we find $\Omega_a/\Omega_d\lesssim 10^{-3}$, with stronger constraints on heavier axions. Lighter axions, however, are allowed and (inflationary) model-independent constraints from the CMB temperature power spectrum and large scale structure are stronger than those implied by tensor modes.

\end{abstract}

\maketitle

\emph{Introduction:} The recent measurement of large angle CMB B-mode polarisation by BICEP2 \cite{bicep}, implying a tensor-to-scalar ratio $r=0.2^{+0.07}_{-0.05}$ has profound implications for our understanding of the initial conditions of the universe \cite{lyth1997}, and points to an inflationary origin for the primordial fluctuations \cite{guth1981,linde1982,albrecht1982}. The inflaton also drives fluctuations in any other fields present in the primordial epoch and so the measurement of $r$, which fixes the inflationary energy scale, can powerfully constrain diverse physics. In this work we will discuss the implications for axion dark matter (DM) in the case that the tensor modes are generated during single-field slow-roll inflation (from now on we simply refer to this as `inflation') by zero-point fluctuations of the graviton. In this work we assume that the measured value of $r$ both holds up to closer scrutiny experimentally, and is taken to be of primordial origin. We relax these assumptions in our closing discussion. We stress that our conclusions are one consequence of taking this measurement at face value, but also that they apply to any detection of $r$.

The scalar amplitude of perturbations generated during inflation is given by \cite{planck2013cosmo}
\be
A_s=\frac{1}{2\epsilon}\left( \frac{H_I}{2\pi M_{pl}}\right)^2=2.19\times 10^{-9}
\ee
where $H_I$ is the Hubble rate during inflation, $\epsilon=-\dot{H}/H^2$ is a slow-roll parameter, and $M_{pl}=1/\sqrt{8 \pi G}=2.4\times 10^{18}$ GeV is the reduced Planck mass. The zero-point fluctuations of the graviton give rise to tensor fluctuations with amplitude
\be
A_T=8\left(\frac{H_I}{2\pi M_{pl}}\right)^2 \, ,
\ee
so that the tensor to scalar ratio is $r=A_T/A_s=16\epsilon$\footnote{The value of $r=0.2$ is in slight tension with current temperature measurements. Increasing the damping in the tail, or violating slow roll helps reduce the tension, albeit in an ad hoc fashion. \cite{bicep,planck2013inflation}. The corrections affect isocurvature amplitudes and $r$ at the percent level and do not substantially alter our conclusions.}. The measured values of $r$ and $A_s$ give:
\be
H_I=1.1\times 10^{14}\text{ GeV}\, .
\label{eqn:hi_tensor}
\ee
It is this high scale of inflation that will give us strong constraints on axion DM.

Axions \cite{pecceiquinn1977,wilczek1978,weinberg1978} were introduced as an extension to the standard model of particle physics in an attempt to dynamically solve the so-called `Strong-\emph{CP} problem' of QCD. The relevant term in the action is the \emph{CP}-violating topological term
\be
S_\theta=\frac{\theta}{32 \pi^2}\int d^4x \epsilon^{\mu\nu\alpha\beta}\text{Tr }G_{\mu\nu}G_{\alpha\beta} \, ,
\ee
where $G_{\mu\nu}$ is the gluon field strength tensor. The $\theta$ term implies the existence of a neutron electric dipole moment, $d_n$. Experimental bounds limit $d_n<2.9\times 10^{-26}$~$e$~cm \cite{baker2006} and imply that $\theta\lesssim 10^{-10}$. The Peccei-Quinn \cite{pecceiquinn1977} (PQ) solution to this is to promote $\theta$ to a dynamical field, the axion \cite{wilczek1978,weinberg1978}, which is the Goldstone boson of a spontaneously broken global $U(1)$ symmetry. At temperatures below the QCD phase transition, QCD instantons lead to a potential and stabilise the axion at the \emph{CP}-conserving value of $\theta=0$. The potential takes the form \cite{gross1981}
\be
V(\phi)=\Lambda^4 (1-\cos\phi/f_a) \, .
\ee
The canonically normalised field is $\phi=f_a\theta$, where $f_a$ is the axion decay constant and gives the scale at which the PQ symmetry is broken. Oscillations about this potential minimum lead to the production of axion DM \cite{turner1983b,turner1983,dine1983,abbott1983,preskill1983,turner1986,berezhiani1992}\footnote{For more details see e.g. Refs.~\cite{Kim:1986ax,book:kolb_and_turner,raffelt2001,sikivie2008}.}. Axions are also generic to string theory \cite{witten1984,witten2006,axiverse2009}, where they and similar particles come under the heading `axion-like particles' (e.g. Ref.~\cite{ringwald2012b}). Along with the QCD axion we will also consider constraints on other axions coming from a measurement of $r$.

Just as the graviton is massless during inflation, leading to the production of the tensor modes, if the axion is massless during inflation (and the PQ symmetry is broken) it acquires isocurvature perturbations \cite{axenides1983,seckel1985}
\be
\sqrt{\langle\delta\phi^2\rangle}=\frac{H_I}{2\pi} \, . 
\label{eqn:iso_fluc}
\ee
Thus high-scale inflation as required in the simplest scenario giving rise to $r$ implies large amplitude isocurvature perturbations \cite{turner1991,fox2004}.

The spectrum of initial axion isocurvature density perturbations generated by Eq.~(\ref{eqn:iso_fluc}) is
\be
 \langle\delta_a^2\rangle=4\left\langle\left(\frac{\delta\phi}{\phi}\right)^2\right\rangle = \frac{(H_I/M_{pl})^2}{\pi^2 (\phi_i/M_{pl})^2} \, .
 \label{eqn:iso_spectrum}
\ee
Given that axions may comprise but a fraction $\Omega_a/\Omega_d$ of the total DM, the isocurvature amplitude is given by 
\be
A_I=\left(\frac{\Omega_a}{\Omega_d}\right)^2\frac{(H_I/M_{pl})^2}{\pi^2 (\phi_i/M_{pl})^2} \, .
\ee
The ratio of power in isocurvature to adiabatic modes is given by:
\be
\frac{A_I}{A_s}=\left(\frac{\Omega_a}{\Omega_d}\right)^2 \frac{8\epsilon}{(\phi_i/M_{pl})^2} \, .
\ee

These isocurvature modes are uncorrelated with the adiabatic mode. The QCD axion is indistinguishable from CDM on cosmological scales, and the Planck collaboration \cite{planck2013inflation} constrains uncorrelated CDM isocurvature to contribute a fraction
\be
\frac{A_I}{A_s}< 0.04 \, .
\label{eqn:planck_iso}
\ee
Given certain assumptions, in particular that the PQ symmetry is broken during inflation and that the QCD axion makes up all of the DM, this implies the limit
\be
H_I\leq 2.4\times 10^9 \text{ GeV} \left( \frac{f_a}{10^{16}\text{ GeV}} \right)^{0.408} \, ,
\ee
which is clearly inconsistent by many orders of magnitude with the value of Eq.~(\ref{eqn:hi_tensor}) implied by the detection of $r$.  

\emph{The QCD Axion:} We now discuss the well known implications of a measurement of $r$ as applied to the QCD axion (e.g. \cite{fox2004,hertzberg2008,5yearWMAP,mack2009a}). For the QCD axion the decay constant is known to be in the window
\be
10^{9}\text{ GeV}\lesssim f_a \lesssim 10^{17}\text{ GeV}\, ,
\ee 
where the lower bound comes from stellar cooling \cite{raffelt2008} and the lesser known upper bound from the spins of stellar mass black holes \cite{arvanitaki2010}.

The homogeneous component of the field $\phi$ evolves according to the Klein-Gordon equation in the expanding universe
\be
\ddot{\phi}+3H\dot{\phi}+V_{,\phi}=0 \, .
\ee
Once Hubble friction is overcome, the field oscillates in its potential minimum, with the energy density scaling as matter, and provides a source of DM in this `vacuum realignment' production. There are various possibilities to set the axion relic density, depending on whether the PQ symmetry is broken or not during inflation.

The relic density due to vacuum realignment is given by
\be
\Omega_a h^2 \sim 2\times 10^4 \left( \frac{f_a}{10^{16}\text{ GeV}}\right)^{7/6}\langle\theta_i^2\rangle \gamma \, ,
\label{eqn:qcd_vac_high_f}
\ee
where angle brackets denote spatial averaging of the short wavelength fluctuations \cite{lyth1992}, $0<\gamma<1$ is a dilution factor if entropy is produced sometime after the QCD phase transition and before nucleosynthesis (for example by decay of a weakly coupled modulus)\footnote{We note that for $10^{15}\text{ GeV}\lesssim f_a \lesssim 10^{17}\text{ GeV}$ there is no exactly known expression for $\Omega_a$ when oscillations begin during the QCD phase transition (e.g. \cite{fox2004,wantz2009}). Also, in order for large entropy production to be possible oscillations must begin in a matter dominated era, giving another slightly different expression (which can be absorbed into $\gamma$) \cite{acharya2010a}.}, and we have dropped the factor $f(\theta_i^2)$ accounting for anharmonic effects for simplicity. 

The PQ symmetry is broken during inflation\footnote{More rigorously the condition is \cite{hertzberg2008} $f_a>\rm{Max}\{ T_{\rm{GH}},T_{\rm{max}}\}$
where $T_{\rm{GH}}$ is the Gibbons-Hawking temperature of de Sitter space during inflation, $T_{\rm{GH}}=H_I/2\pi$ \cite{gibbons1977,bunch1978} and $T_{\rm{max}}$ is the maximum thermalisation temperature after inflation, $T_{\rm{max}}=\gamma_{\rm{eff}} E_I$ ($\gamma_{\rm{eff}}$ is an efficiency parameter and $E_I=3^{1/4}\sqrt{M_{pl}H_I}$).} if $f_a>H_I/2\pi$ and then the homogeneous component of $\theta$ is a free parameter in each horizon volume. Even in the simplest case where $\langle\theta_i^2\rangle\sim \bar{\theta}_i^2$, then  for large $f_a\sim 10^{16}\text{ GeV}$ Eq.~(\ref{eqn:qcd_vac_high_f}) already implies a modest level of fine tuning to $\theta_i\sim 10^{-2}$ if the axion is not to overclose the universe, $\rho_a>\rho_{\rm crit}$, where $\rho_{\rm crit}$ is the critical density for flatness. However, this fine tuning is easy to accommodate in the so-called `anthropic axion window' \cite{hertzberg2008}. 

Combining Eqs.~(\ref{eqn:planck_iso}), and (\ref{eqn:qcd_vac_high_f}) with the measured value of $r$ and setting $\Omega_d h^2=0.119$ \cite{planck2013cosmo}, the tensor and isocurvature constraints put an upper limit on the axion DM fraction of
\be
\frac{\Omega_{a,{\rm QCD}}}{\Omega_d}\lesssim \frac{4\times 10^{-12}}{\gamma} \left(\frac{f_a}{10^{16}\text{ GeV}} \right)^{5/6}\left(\frac{0.2}{r} \right)\left(\frac{\Omega_d h^2}{0.119} \right) \, .
\ee
This constraint essentially rules out the high-$f_a$ QCD axion as a DM candidate, showing the far reaching implications of the measurement of $r$. Barring an impossibly huge \cite{fox2004} dilution of axion energy density, $\gamma\ll 1$, this small abundance gives an upper limit on the QCD axion effective initial misalignment angle
\be
\langle\theta_i^2\rangle\lesssim \frac{2\times 10^{-17}}{\gamma^2} \left(\frac{f_a}{10^{16}\text{ GeV}} \right)^{-1/3}\left(\frac{\Omega_d h^2}{0.119} \right)^2\left(\frac{0.2}{r} \right) \, .
\label{eqn:theta_qcd_dm}
\ee

In low $f_a$ models the axion does not acquire isocurvature perturbations since the field is not established when the PQ symmetry is unbroken. Therefore with low-$f_a$ there is no additional constraint on axions derived from combining the measurement of $r$ with the bound on $A_I/A_s$, other than setting the scale for this scenario. When the PQ symmetry is broken after inflation, the axion field varies on cosmologically small scales with average $\langle\theta^2\rangle=\pi^2/3$, which should be used in Eq.~(\ref{eqn:qcd_vac_high_f}) to compute the relic abundance. The requirement of not overproducing DM, $\Omega_a h^2<0.119$, then limits the maximum value of $f_a$ to $f_a<1.2\times 10^{11} \chi^{-6/7}\text{ GeV}$ \cite{hertzberg2008} where $\chi$ can vary by an order of magnitude or more and accounts for theoretical uncertainties (including production from string decay)\footnote{See e.g. Ref.~\cite{wantz2009} where it is argued that the value of $f_a$ assuming no string contribution, $\chi=1$, still gives a useful benchmark for the excluded masses.}. For low $f_a$ there are relics of the PQ transition no longer diluted by inflation \cite{book:kolb_and_turner}. While domain walls are problematic, string decay can be the dominant source of axion DM in this scenario. The case of low $f_a$ axions has been discussed extensively elsewhere, and we discuss them no further here.

\begin{figure}[htbp!]
\includegraphics[scale=0.41]{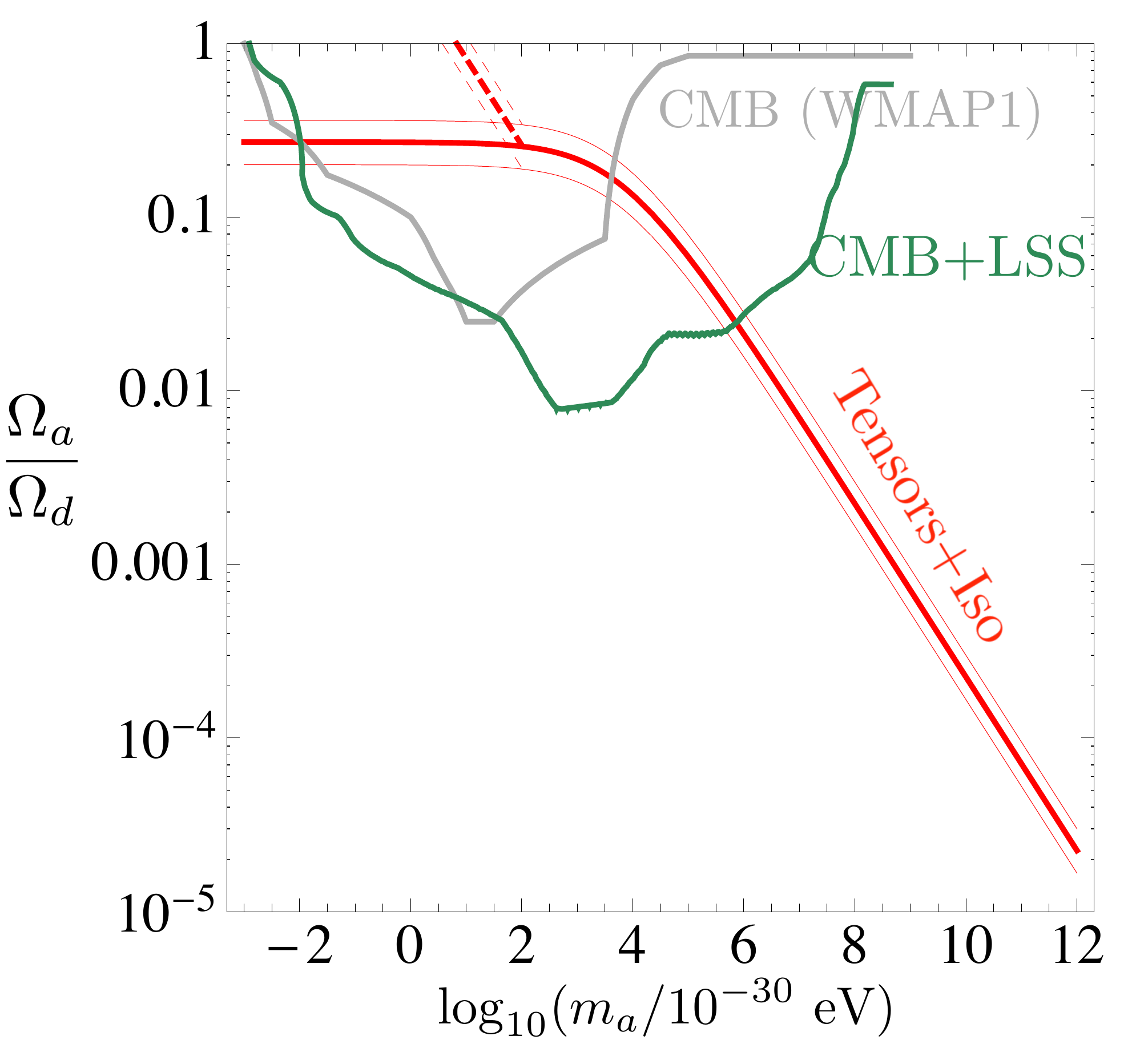}\\[0.0in] 
\caption{Constraints in axion parameter space: regions below curves are allowed. The solid red line shows the result of the present work which constrains axions using the measured value of $r=0.2$ ($^{+0.07}_{-0.05}$ shown in thin lines) and the Planck constraint on axion isocurvature, $A_I/A_s<0.04$. The dashed red line approximates the loosening of this constraint due to suppression of the axion isocuvature power when $m_a<H_{\rm eq}$. We also show the 95\% exclusion contours of Ref.~\cite{amendola2005} from CMB (WMAP1) and CMB+Lyman-alpha forest power spectra, which are significantly stronger than the tensor/isocurvature constraint for intermediate mass axions, and are independent of the inflationary model.}
\label{fig:axions_tensors}
\end{figure}

\emph{Ultra-light Axions:} In this section we further develop the ideas presented in Ref.~\cite{marsh2013} and show an \emph{estimate} of the combined constraints on axion parameter space from isocurvature, a confirmed detection of $r$, and other cosmological constraints of Ref.~\cite{amendola2005}. 

Ultra-light axions are motivated by string theory considerations, with the mass scaling exponentially with the moduli \cite{axiverse2009}, or simply by a Jeffreys prior on this unknown parameter. They differ from the QCD axion in that they need not couple to QCD, or indeed the standard model. For such a generic axion the temperature dependence of the mass cannot be known, as the masses arise from non-perturbative effects in hidden sectors. As long as the mass has reached its zero-temperature value by the time oscillations begin, the relic abundance due to vacuum realignment is given by
\be
\Omega_a\approx\frac{a_{\rm osc}^3}{6 H_0^2}m_a^2 \left\langle\left(\frac{\phi_i}{M_{pl}}\right)^2\right\rangle \, ,
\label{eqn:alp_abundance}
\ee
where $a_{\rm osc}$ is the scale factor defined by $3H(a_{\rm osc})=m_a$ when oscillations begin: it can be approximated by using the Friedmann equation and assuming an instantaneous transition in the axion equation of state from $w_a=-1$ to $w_a=0$ at $a_{\rm osc}$. When $m_a\lesssim 10^{-18}\text{ eV}$ the relic abundance cannot be significant unless $f_a\gtrsim 10^{16}\text{ GeV}>H_I$ and therefore in what follows we consider only the case where the PQ symmetry is broken during inflation\footnote{For a single axion this is true, but for many axions, as in the axiverse \cite{axiverse2009}, an N-flation type scenario for DM could be relevant.}.

Pressure perturbations in axions can be described using a scale-dependent sound speed, leading to a Jeans scale below which density perturbations are suppressed \cite{hu2000,amendola2005,axiverse2009,marsh2010,park2012,marsh2013b}. When the mass is in the range $10^{-33}\text{ eV}\lesssim m_a\lesssim 10^{-18}\text{ eV}$ this scale can be astrophysical or cosmological in size and therefore can be constrained using the CMB power spectrum and large-scale structure (LSS) measurements \cite{amendola2005,marsh2011b,marsh_etal_inprep}. The size of the effect is fixed by the fraction of DM in axions, $\Omega_a/\Omega_d$, and so constraints are presented in the $(m_a,\Omega_a/\Omega_d)$ plane. Constraints from the CMB are particularly strong for $m_a\lesssim H_{\rm eq}\sim 10^{-28}\text{ eV}$ where the axions roll in their potential after equality, shifting equality and giving rise to an Integrated Sachs-Wolfe (SW) effect from the evolving gravitational potential \cite{marsh2011b}.

Light axions also carry their own isocurvature perturbations \cite{marsh2013}, with the spectrum Eq.~(\ref{eqn:iso_spectrum}). Fixing the initial field displacement in terms of the DM contribution from Eq.~(\ref{eqn:alp_abundance}) allows us to place a constraint across the $(m_a,\Omega_a/\Omega_d)$ plane given by the measured value of $r$ and the Planck constraint on $A_I/A_s$. The measured value of $r$ restricts the allowed values of $\Omega_a$ to be small. We show this constraint with the solid red line on Fig.~\ref{fig:axions_tensors}, along with the CMB (WMAP1) and LSS (Lyman-alpha forest) constraints of Ref.~\cite{amendola2005}. Regions below curves are allowed.

The Planck constraints on axion isocurvature apply only to the case where the axions are indistinguishable from CDM, however the suppression of power due to axion pressure shows up also in the isocurvature power for low masses \cite{marsh2013} and the Planck constraints cannot be applied. Work on constraining this mode is ongoing \cite{marsh_etal_inprep}. The CMB isocurvature constraint is driven by the SW plateau. As the axionic Jeans scale crosses into the SW plateau at low mass and suppresses the isocurvature transfer function \cite{marsh2013}, the signal-to-noise $SNR\propto 1/l_{\rm max}$, where $l_{\rm max}\sim l_{\rm Jeans}\sim \sqrt{m_{\rm a}}$. Therefore we estimate that the isocurvature limit is given by $(A_I/A_s)^{\rm max}\propto (A_I/A_s)^{\rm max}_{\rm old}\times \sqrt{10^{-28}~{\rm eV}/m_{\rm a}}$. This estimate is used to obtain the dashed line in Fig.~\ref{fig:axions_tensors}.

Fig.~\ref{fig:axions_tensors} shows the huge power of the measurement of $r$ to constrain axions, giving $\Omega_a/\Omega_d<10^{-3}$ for $m_a\gtrsim 10^{-22}\text{ eV}$, far beyond the reach even of the Lyman-alpha forest constraints. For $m_a\lesssim 10^{-24}\text{ eV}$, however, the constraints from the CMB temperature and E-mode polarisation and LSS (WMAP1 and SDSS \cite{amendola2005}, Planck and WiggleZ in preparation \cite{marsh_etal_inprep}) are stronger than the tensor/isocurvature constraint, and are independent of the inflationary interpretation of BICEP2. 

\emph{Ruling out axions:} Spatial averaging of short wavelength modes gives rise to an irreducible back-reaction contribution to $\langle\phi^2\rangle$ and thus $\Omega_a$. If the required small values cannot be obtained, the corresponding axion is ruled out. Specifically
\be
\langle\phi^2\rangle = \bar{\phi}^2 + \sigma_\phi^2 = \bar{\phi}^2+\langle\delta\phi^2\rangle\, .
\ee
The mean homogeneous value, $\bar{\phi}$, can be tuned or dynamically made arbitrarily small (e.g. via coupling to a tracking field \cite{marsh2011,marsh2012}); fixing $\bar{\phi}=0$ gives the irreducible contribution to $\Omega_a$ from fluctuations. Plugging the variance into Eq.~(\ref{eqn:theta_qcd_dm}) we find that the QCD axion with $f_a>H_I/2 \pi$ is totally ruled out \cite{fox2004} (unless also $f_a\gg M_{pl}$), further taking the low $f_a$ value above this rules out $m_a\lesssim 52\chi^{6/7}\,\mu\text{eV}$. Applying this to the ultra-light axion abundance in Eq.~(\ref{eqn:alp_abundance}) we find that $\Omega_a/\Omega_d<10^{-7}$ over the entire range of masses we consider, which is always below the amount necessary to satisfy the tensor plus isocurvature constraint, and thus no ultra-light axions are completely excluded. This is because order Planckian field displacements are necessary for non-negligible abundance in ultra-light axions, while $H_I<M_{pl}$ sources the fluctuation contribution.

\emph{Discussion:} We have considered the implications of the BICEP2 detection of $r$ on axion DM. In the simplest inflation models $r=0.2$ \cite{bicep} implies $H_I=1.1\times10^{14}\text{ GeV}$. Axions with $f_a>H_I/2\pi$ acquire isocurvature perturbations and are constrained strongly by the Planck bound $A_I/A_s<0.04$. All such high $f_a$ QCD axions are ruled out. Even if they can exist (by somehow suppressing the fluctuation contribution to the abundance), evading isocurvature bounds will require searches for them to be independent of the DM abundance \cite{arvanitaki2014}. In the general, non-QCD, case low $f_a<H_I/2\pi$ axions \cite{cicoli2012c} are unaffected by the tensor bound. High $f_a$ axions \cite{axiverse2009,acharya2010a} are strongly constrained, although for $m_a\lesssim 10^{-28}\text{ eV}$ suppression of power in the isocurvature mode can loosen constraints \cite{marsh2013}. One may consider the high-$f_a$ ultra-light axions `guilty by association' to the QCD axion, but this is a model-dependent statement and axion hierarchies are certainly possible \cite{Kim:2004rp} and indeed desirable if the inflaton is also an axion, as many high $H_I$ models demand.

There are in principle (at least) five ways around the isocurvature bounds. The first is to produce gravitational waves during inflation giving $r=0.2$ while keeping $H_I$ low \cite{senatore2011,cook2011}. Secondly, entropy production after the QCD phase transition can dilute the QCD axion abundance. This is possible in models with light moduli and low temperature reheating (e.g. \cite{iliesiu2013} and references therein). Light axions oscillate after nucleosynthesis and cannot be diluted by such effects. Thirdly, if the axions are massive during inflation they acquire no isocurvature, although a shift symmetry protects axion masses. Fourthly, non-trivial axion dynamics during inflation suppressing isocurvature are possible e.g. via non-minimal coupling to gravity \cite{folkerts2013} or coupling the inflation directly to the sector providing non-perturbative effects, e.g. the QCD coupling \cite{dvali1995,jeong2013b}. Such couplings may alter the adiabatic spectrum and produce observable signatures through production of primordial black holes. Finally coupling a light ($m_a\lesssim 10^{-28}\text{ eV}$) axion to $\vec{E}\cdot\vec{B}$ of electromagnetism could induce `cosmological birefringence' \cite{carroll1990} leading to production of B-modes that are not sourced by gravitational waves \cite{pospelov2008,axiverse2009}. This possibility will be easy to distinguish from tensor and lensing B-modes by its distinctive oscillatory character at high $\ell$, measurable for example by SPTPol and ACTPol.

Other cosmological constraints on axions are more powerful than the tensor/isocurvature bound for light masses $m_a\lesssim 10^{-24}$ \cite{amendola2005,marsh_etal_inprep}. We are exploring this mass range with a careful search of parameter space using nested sampling  \cite{marsh2013}. Isocurvature constraints will improve in the future \cite{hamann2009}, as will constraints on $\Omega_a/\Omega_d$ \cite{marsh2011b}, both of which could allow for a detection consistent with the tensor bound \cite{marsh2013}. In the regime $m_a\gtrsim 10^{-24}\text{ eV}$ the tensor bound is stronger than current cosmological bounds on $\Omega_a$. However, in this regime axions can play a role in resolving issues with galaxy formation if they are dominant in DM \cite{marsh2013b}. Future weak lensing surveys will cut into this regime \cite{lensing_inprep} and surpass the indirect tensor bound. If these axions are necessary/detected in large scale structure this would imply either contradiction with the tensor bound, or other new physics during inflation. The same is true for direct detection of a high $f_a$ QCD axion DM \cite{budker2013}.

{\it Note added in proof:} The related paper Ref.~\cite{visinelli2014} referring to the QCD axion has also recently appeared.

\vspace{-0.25in}
% ------------------------ ACKNOWLEDGEMENTS ----------------------------------
\begin{acknowledgments}
\vspace{-0.1in}
We are especially grateful to the anonymous referee, whose suggestions greatly improved the manuscript. We are grateful to Luca Amendola for providing us with the contour constraints of Ref.~\cite{amendola2005}, and to Asimina Arvanitaki, Piyush Kumar and Maxim Pospelov for discussions. PGF acknowledges support from STFC, BIPAC and the Oxford Martin School. DG is funded at the University of Chicago by a National Science Foundation Astronomy and Astrophysics Postdoctoral Fellowship under Award NO. AST-1302856. Research at Perimeter Institute is supported by the Government of Canada through Industry Canada and by the Province of Ontario through the Ministry of Research and Innovation.
\end{acknowledgments}

% \renewcommand{\theequation}{A\arabic{equation}}
  % redefine the command that creates the equation no.
 % \setcounter{equation}{0} 

% ---------------------- BIBLIOGRAPHY -----------------------------------------

\bibliographystyle{h-physrev3}
\bibliography{doddyoxford}
\end{document}